\documentclass[aps,prb,twocolumn,amssymb,amsmath,superscriptaddress]{revtex4-1}
\usepackage{bm}
\usepackage{times}
\usepackage{graphicx}
\usepackage{color}
\usepackage{dcolumn}
\usepackage[colorlinks=true, letterpaper=true, pdfstartview=FitV, linkcolor=blue, citecolor=blue, urlcolor=blue]{hyperref}
\usepackage{appendix}

\begin{document}
\title{Quantized valley source in bilayer graphene via adiabatic quantum pumping}

\author{Kai-Tong Wang}
\affiliation{School of Physics and Engineering, Henan University of Science and Technology, Luoyang 471023, China}
\affiliation{College of Physics and Optoelectronic Engineering, Shenzhen University, Shenzhen 518060, China}

\author{Lei Zhang}
\affiliation{State Key Laboratory of Quantum Optics Technologies and Devices, Institute of Laser Spectroscopy, Shanxi University, Taiyuan 030006, China}
\affiliation{Collaborative Innovation Center of Extreme Optics, Shanxi University, Taiyuan 030006, China}

\author{Bin Wang}
\affiliation{College of Physics and Optoelectronic Engineering, Shenzhen University, Shenzhen 518060, China}

\author{Yunjin Yu}
\affiliation{College of Physics and Optoelectronic Engineering, Shenzhen University, Shenzhen 518060, China}

\author{Fuming Xu}
\email[]{xufuming@szu.edu.cn}
\affiliation{College of Physics and Optoelectronic Engineering, Shenzhen University, Shenzhen 518060, China}
\affiliation{Quantum Science Center of Guangdong-Hongkong-Macao Greater Bay Area, Shenzhen 518045, China}


\begin{abstract}
We propose a quantized valley source based on adiabatic quantum pumping in bilayer graphene in the presence of both quantum spin Hall and quantum valley Hall effects. By periodically modulating the gate voltages of two quantum dots coupled to a bilayer graphene nanoribbon, the helical edge states enable the generation of a pure valley current. Depending on the geometry of quantum dots, the number of emitted valley pairs either forms quantized plateaus or varies continuously with the pumping amplitude. The quantized plateau structure can be tuned by the static gate voltage and by the coupling between the quantum dots and the ribbon. Quantized valley pairs are robust against Anderson disorder applied in the quantum dots and persist for both zigzag and armchair quantum-dot boundaries. These results demonstrate that adiabatic quantum pumping provides a route toward electrically controllable quantized valley sources.
\end{abstract}
\maketitle

\section{Introduction}

Many crystalline materials host symmetry-related extrema in their band structures, known as valleys. These degenerate but inequivalent valleys have attracted considerable interest because of their potential for encoding information. Analogous to spin, the valley index can serve as an information carrier, giving rise to the field of valleytronics\cite{LundebergHarnessing,Schaibley2016Valleytronics,HuMoire,StevenValleytronics,Yanping2019Valleytronics,ChenValley,Zhi2020Valley}. A central operation in valleytronics is the generation of valley polarization. Proposed approaches include optical selection rules\cite{MakControl,Zhang14Generation,Ye2016Optical,Yang24Photocurrent,Sharma24Direct,Gill25Creation}, structural engineering\cite{GunlyckeGraphene,Fuming2016Transmission}, magnetic fields\cite{LiValley,Park2019Magnetoelectrically,CaiMagnetic}, and nonmagnetic disorder\cite{AnRealization}. Recent transport proposals further exploit valley-selective Klein tunneling in graphene superlattices\cite{An20Klein}, electrically tunable Rashba backscattering\cite{Zhang23Electrical}, and valley-resolved Fano resonances\cite{Mu22Fano}. Other device concepts employ non-Hermitian valley filtering in moir\'{e} bilayer graphene\cite{Shao24NonHermitian} and spin-valley-mismatched ferromagnetic tunnel junctions\cite{Yan25Giant}. Practical valleytronic applications therefore require controllable sources of valley-polarized carriers\cite{AngValley}.

An ideal single-electron source emits or reabsorbs one electron during each driving cycle\cite{FeveAn,MoskaletsQuantized}, providing a building block for quantum electronics in ballistic conductors. Related proposals have considered sources of spin-polarized carriers\cite{Gurzhi2003Spin}, including a coherent single-spin source that generates a pure quantized spin current under periodic gate modulation\cite{Xing2014Coherent}. Spin-valley locking has been studied in two-dimensional (2D) transition-metal dichalcogenides\cite{XiaoCoupled,KhaniGate}, silicene\cite{EzawaSpin,XiongSpin}, and Dirac semimetal ${\mathrm{BaMnSb}}_{2}$\cite{LiuSpin2021}, enabling electrical control of the valley degree of freedom\cite{Ominato2020,Yang2020,wang2023}. Related schemes include electrical switching of spin and valley polarization in graphene multilayers\cite{Wang24Electrical}, valleytronic setups\cite{ZhuQuantum,TaoTwo,TaoValley}, and valley-based quantum information processing\cite{WuSpin,GImpurity,PawlowskiSpin}. These developments motivate the construction of a controllable valley source.

Topological insulators (TIs) provide a natural platform for extending controlled carrier emission to coupled spin and valley transport\cite{BernevigQuantum,KonigQuantum,ZhangTopological,ChenExperimental,Wu2017Observation,VergnioryA}. In 2D, time-reversal-protected helical edge states of the quantum spin Hall (QSH) effect are robust against nonmagnetic disorder. 2D honeycomb materials that couple spin and valley degrees of freedom exhibit the QSH effect\cite{LiuQuantum}, the valley Hall effect\cite{XiaoValley}, the valley Seebeck effect\cite{YuValley,ZhangInfluence}, and the valley-polarized quantum anomalous Hall effect\cite{Pan2014Valley,Pan2015Valley,Zhai2020}. The connection between a bulk valley Hall current and observable edge accumulation can, however, depend on valley-density nonconservation and edge termination\cite{Kazantsev24Nonconservation}. In gated bilayer graphene, Rashba spin-orbit coupling can drive a transition between two topologically distinct phases\cite{QiaoTwo}, producing a phase in which the QSH and quantum valley Hall effects coexist. This coexistence provides a promising platform for controlling the valley degree of freedom.

Recent work on graphene/hBN heterostructures has shown that synchronized time-dependent gate voltages can dynamically control a periodically modulated, valley-polarized current\cite{Belayadi25Dynamic}. In this work, we use a similar gate-controlled pumping scheme in bilayer graphene to generate a pure valley current with quantized valley pairs. This mechanism is realized by combining adiabatic parametric pumping\cite{SwitkesAn,WeiResonance} with topological edge-state selectivity. Specifically, we periodically modulate the gate voltages applied to two quantum dots (QDs) coupled to a bilayer graphene ribbon. Depending on the phase difference between the two gate voltages, the device generates either a pure charge current or a pure valley current. For the out-of-phase configuration($\phi=\pi$), slender QDs emit one or more quantized valley pairs. In constrast, for flattened QDs, the number of pumped valley pairs varies approximately linearly with the pumping amplitude. We find that quantization is governed primarily by (i) the discrete QD energy levels and (ii) the coupling between the QDs and the bilayer graphene ribbon. The quantized plateau structure can therefore be controlled through the static gate voltage, QD geometry, and QD-ribbon coupling. Quantized valley response also persists for armchair QD boundaries and is robust against disorder confined to the QDs.

The rest of this paper is organized as follows. In Sec.~\uppercase\expandafter{\romannumeral2}, we introduce the tight-binding Hamiltonian for bilayer graphene and the formalism for calculating the pumped valley response. Sec.~\uppercase\expandafter{\romannumeral3} presents the numerical results and discussion. A summary is given in Section~\uppercase\expandafter{\romannumeral4}.

\section{Model and Formalism}

\begin{figure}
\centering
\includegraphics[width=\columnwidth]{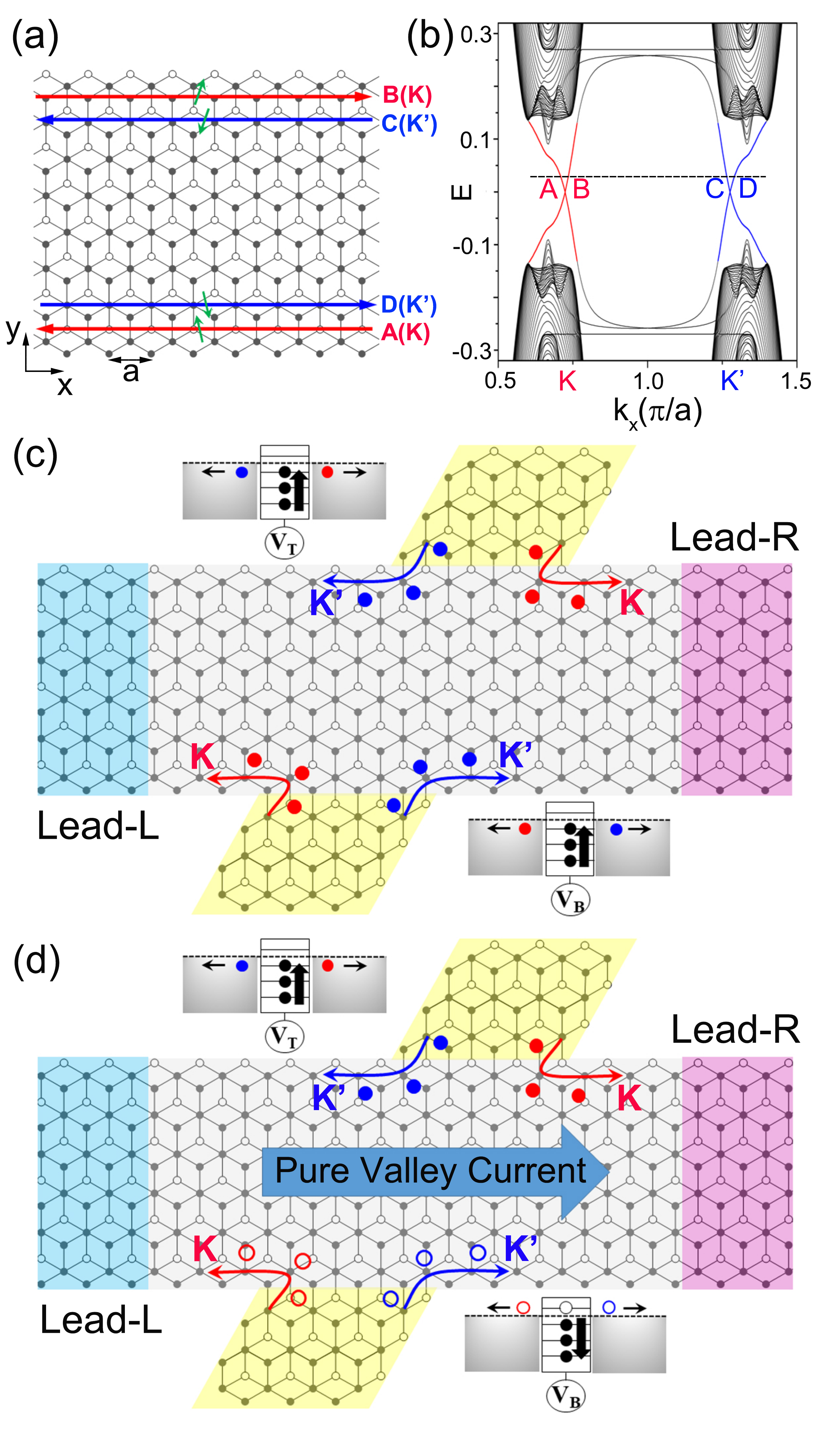}
\caption{(a) Schematic of an AB-stacked bilayer graphene ribbon with zigzag boundaries, showing the propagation of edge states associated with the QSH and QVH effects. Sites in the upper and lower layers are represented by black and white circles, respectively. (b) Band structure of the zigzag ribbon along the $x$ direction. Labels A--D identify the edge states depicted in panel (a). (c) and (d) Schematics of the valley-resolved pumping processes into the left and right leads in half a cycle. The device generates a pure charge current for $\phi=0$ and a pure valley current for $\phi=\pi$, where $\phi$ is the phase difference between the gate voltages $V_T$ and $V_B$. The corresponding voltage configurations are shown in the insets.}\label{fig1}
\end{figure}

In the presence of an interlayer gate potential and Rashba spin-orbit (SO) coupling, AB-stacked bilayer graphene can enter the QSH regime. Its helical edge states distinguish both spin and valley, thereby enabling valley filtering. As illustrated in Fig.~\ref{fig1}(a), the QSH and quantum valley Hall (QVH) effects coexist in this regime.

For AB-stacked bilayer graphene subject to Rashba SO coupling and an interlayer potential difference, the tight-binding Hamiltonian is\cite{QiaoTwo}
\begin{eqnarray}
H_{\rm BLG}=&H_{\rm SLG}^T +H_{\rm SLG}^B +t_{\bot}\sum_{i\in T,j\in B,\alpha}(d_{i\alpha}^{\dag} d_{j\alpha}+d_{j\alpha}^{\dag}d_{i\alpha}) \nonumber \\
           &+U\sum_{i\in T,\alpha}d_{i\alpha}^{\dag} d_{i\alpha} -U\sum_{i\in B,\alpha}d_{i\alpha}^{\dag} d_{i\alpha}, \label{eq1}
\end{eqnarray}
with
\begin{equation}
H_{\rm SLG} = -t\sum_{<ij>\alpha} d_{i\alpha}^{\dag} d_{j\alpha} + it_{R}\sum_{<ij>\alpha\beta} (\mathbf{s}_{\alpha\beta} \times \mathbf{r}_{ij})_{z} d_{i\alpha}^{\dag} d_{j\beta},\label{eq2}
\end{equation}
Here, $H_{\rm SLG}^{T}$ and $H_{\rm SLG}^{B}$ are the single-layer Hamiltonians of the top ($T$) and bottom ($B$) graphene layers, respectively. The operators $d_{i\alpha}^{\dag}$ and $d_{i\alpha}$ create and annihilate an electron with spin $\alpha$ at site $i$. The parameters $t$ and $t_R$ denote the nearest-neighbor hopping in monolayer graphene and the Rashba SO coupling strength, respectively. In addition, $\mathbf{s}$ is the vector of spin Pauli matrices, and $\mathbf{r}_{ij}$ is the lattice vector between sites $i$ and $j$. In Eq.~(\ref{eq1}), $t_{\bot}$ denotes the interlayer hopping, and the final two terms describe the interlayer potential difference $2U$.

When the Fermi energy lies inside the band gap in the QSH regime, the valley indices are locked to the time-reversal-protected helical edge states shown in Fig.~\ref{fig1}(b). This locking gives rise to topological edge-state selectivity: the valley index of a propagating mode is correlated with its edge and direction of motion. To use this selectivity in a quantized valley source, two QDs are coupled to the upper and lower edges of a bilayer graphene ribbon, as shown in Figs.~\ref{fig1}(c) and \ref{fig1}(d). Time-dependent gate voltages $V_T(t)$ and $V_B(t)$ applied to the top and bottom QDs (yellow regions) modulate the QD energy levels and pump valley-polarized carriers into the leads.

We first consider valley transport through the top QD. In the first half of a cycle of $V_T(t)$, electrons of a given valley are emitted from the QD into the right lead. In the second half of the cycle, holes of the same valley are emitted, ensuring particle conservation over a complete cycle. Electrons and holes of the opposite valley undergo an analogous process toward the left lead. The bottom QD operates similarly under $V_B(t)$, emitting electrons and holes into the left and right leads over one complete cycle. We assume equal modulation amplitudes for the two gate voltages. When $V_T(t)$ and $V_B(t)$ are in phase ($\phi=0$), two electrons with opposite valley indices are emitted into the left and right leads in the same half of the cycle, producing a quantized charge current [see the inset of Fig.~\ref{fig1}(c)]. When the gate voltages are out of phase ($\phi=\pi$), an electron and a hole with opposite valley indices are emitted from the top and bottom QDs, respectively [see the inset of Fig.~\ref{fig1}(d)]. The net charge current then vanishes, whereas a quantized valley current flows from the left lead to the right lead in that half of the cycle, forming a quantized valley emitter.

To calculate the valley-polarized current in the adiabatic regime, we consider the time-dependent pumping potential generated by the gate voltages. The system remains close to equilibrium as the pumping potentials vary slowly. At zero temperature, the pumped current into lead $m$ is calculated as\cite{WeiCarbon,WangCurrent,XuNumerical}
\begin{equation}
I_{m} = -\frac{q}{2\pi}\int {\rm Tr} [\Gamma_{m}G^{r} \frac {d\mathcal{U}}{dt}G^{a}]dt, \label{eq3}
\end{equation}
where $m=L,R$ labels the lead. The retarded Green's function is defined as ${G}^{r}(E,\mathcal{U}(t))={G}^{a}(E,\mathcal{U}(t))^{\dag}=[EI-H-\Sigma_{L}^{r}(E)-\Sigma_{R}^{r}(E)-\mathcal{U}(t)\Delta]^{-1}$, where $H$ is the Hamiltonian of the scattering region and $\Sigma_L^r$ ($\Sigma_R^r$) is the retarded self-energy of the left (right) lead. The quantity $\mathcal{U}$ represents the gate-induced potential on the top and bottom QDs, and $\Delta=\Delta_t+\Delta_b$ specifies the corresponding spatial profile.

To resolve the spin and valley contributions, we construct an effective linewidth function by selecting the corresponding propagating modes\cite{Fuming2016Transmission,XingTopological,WangRelation}
\begin{equation}
\overline{\Gamma}_{m,\alpha,\nu} = \Lambda_{\alpha,\nu} \widetilde{\Gamma}_{m,\alpha,\nu} \Lambda_{\alpha,\nu}^{\dag}, \label{eq4}
\end{equation}
with
\begin{equation*}
\widetilde{\Gamma}_{m} = \rm{U^{\dag}} \Gamma_m \rm{U}.
\end{equation*}
Here, $\alpha$ and $\nu$ denote the spin and valley indices, respectively, and $\Lambda_{\alpha,\nu}$ selects the eigenmodes of $\Gamma_m$ with the specified spin and valley. The unitary transformation matrix $\rm U$ is constructed from the eigenmodes of $\Gamma_m$. The number of valley-polarized electrons is obtained by integrating Eq.~(\ref{eq3}) over half a cycle. Substituting Eq.~(\ref{eq4}) into Eq.~(\ref{eq3}) gives
\begin{equation}
N_{m,\alpha,\nu}=-\frac{1}{2\pi}\int_{0}^{T/2} {\rm Tr} [\overline{\Gamma}_{m,\alpha,\nu} G^{r} \frac{d\mathcal{U}}{dt}G^{a}] dt,\label{eq5}
\end{equation}
where $\alpha=\pm1$ labels the spin and $\nu=K,K'$ denotes the valley. For pure valley pumping ($\phi=\pi$), a valley pair consists of an electron in one valley and a hole in the opposite valley emitted into the same lead in half a cycle. Both electron and hole contribute one member of the pair, and hence the number of valley pairs is given by $|N_{m,\alpha,\nu}|$. Since current is conserved, in the following, we report this pair number in the right lead.

\section{Numerical Results and Discussion}
\begin{figure}
\centering
\includegraphics[width=\columnwidth]{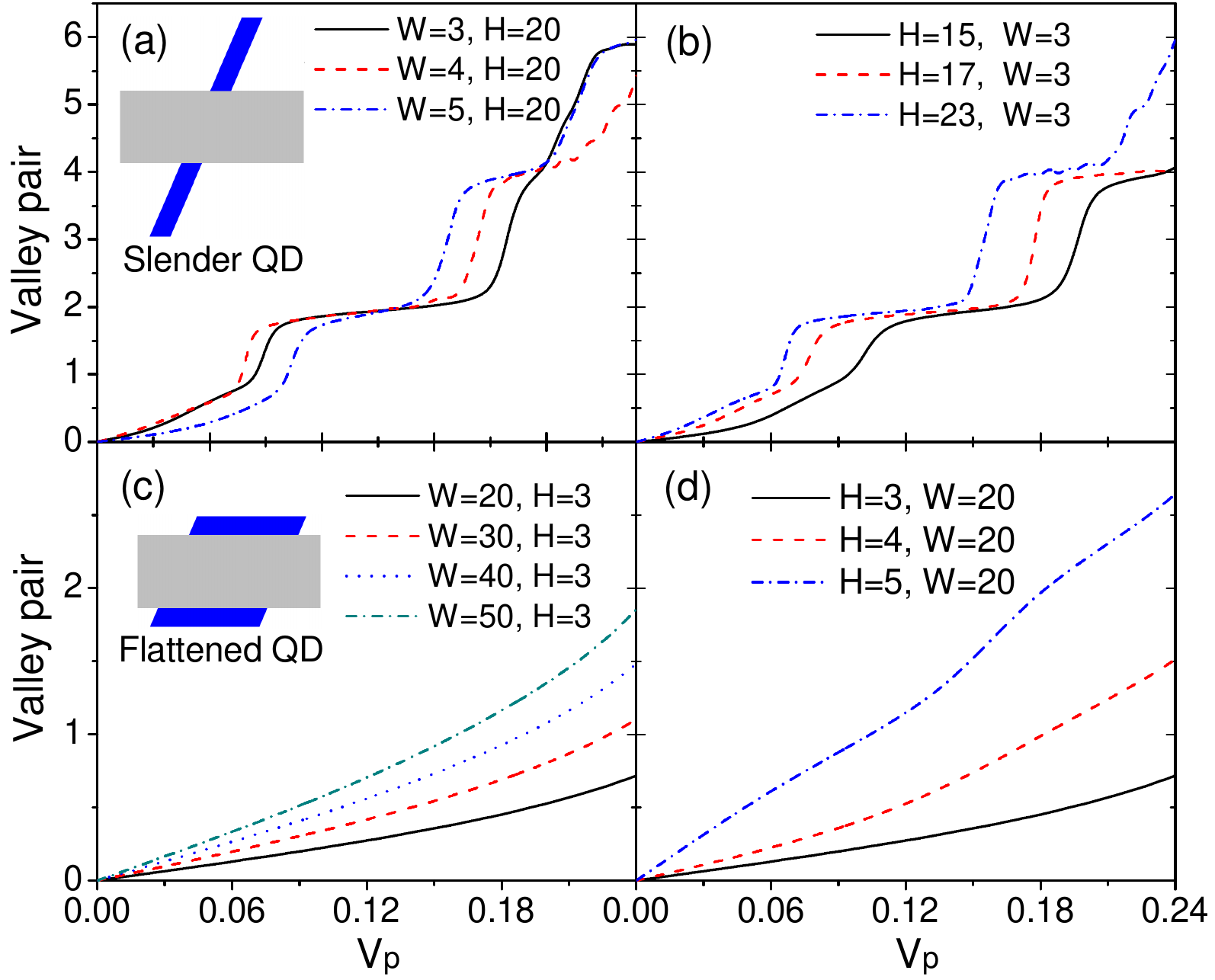}
\caption{Number of valley pairs pumped in half a cycle as a function of the pumping amplitude $V_p$. (a) and (b) Slender QDs coupled to the central ribbon. (c) and (d) Flattened QDs coupled to the ribbon. The parameters $W$ and $H$ denote the width and height of each QD, respectively, and the central ribbon has a width of $80$. The insets illustrate the two device geometries; the blue regions indicate the QDs to which the pumping potentials are applied. The static potential is $V_0=0$.}\label{fig2}
\end{figure}

In the numerical calculations, the bilayer-graphene parameters are $t=2.7~\rm eV$, $t_{\bot}=0.39~\rm eV$, $t_R=0.2~\rm eV$\cite{Yu2008Rashba}, and $U=0.3~\rm eV$. For these parameters, the system supports both QSH and QVH behavior. The Fermi energy in both leads is fixed at $E_f=0~\rm eV$. Throughout this work, energies are expressed in electron volts and lengths in units of the lattice constant $a$. The energy levels of an isolated QD are distributed symmetrically about zero energy. The gate voltages are written as $V_{T/B}(t)=V_0\pm V_p\cos(\omega t)$, where $V_0$ and $V_p$ are the static potential and pumping amplitude, respectively. As a gate voltage sweeps across the discrete QD levels, an instantaneous valley current is emitted into the leads.

We consider two device geometries in which either slender or flattened QDs are coupled to the bilayer graphene ribbon, as shown in the insets of Fig.~\ref{fig2}. The full system has zigzag boundaries along the transport direction to preserve the topological edge states. Figure~\ref{fig2} shows the number of pumped valley pairs as a function of $V_p$ for several QD sizes. A comparison of Figs.~\ref{fig2}(a) and \ref{fig2}(c) reveals distinct pumping behavior in the two geometries. For slender QDs (SQDs), Fig.~\ref{fig2}(a) displays quantized plateaus, demonstrating the emission of an integer number of valley pairs in half a cycle. Increasing the height-to-width ratio of the SQDs produces longer and better-defined plateaus. For flattened QDs (FQDs), by contrast, the number of valley pairs increases approximately linearly with $V_p$, as shown in Figs.~\ref{fig2}(c) and \ref{fig2}(d). At a fixed $V_p$, a larger FQD emits more valley pairs. These results identify the SQD geometry as the more suitable configuration for a quantized valley source.

\begin{figure}
\centering
\includegraphics[width=\columnwidth]{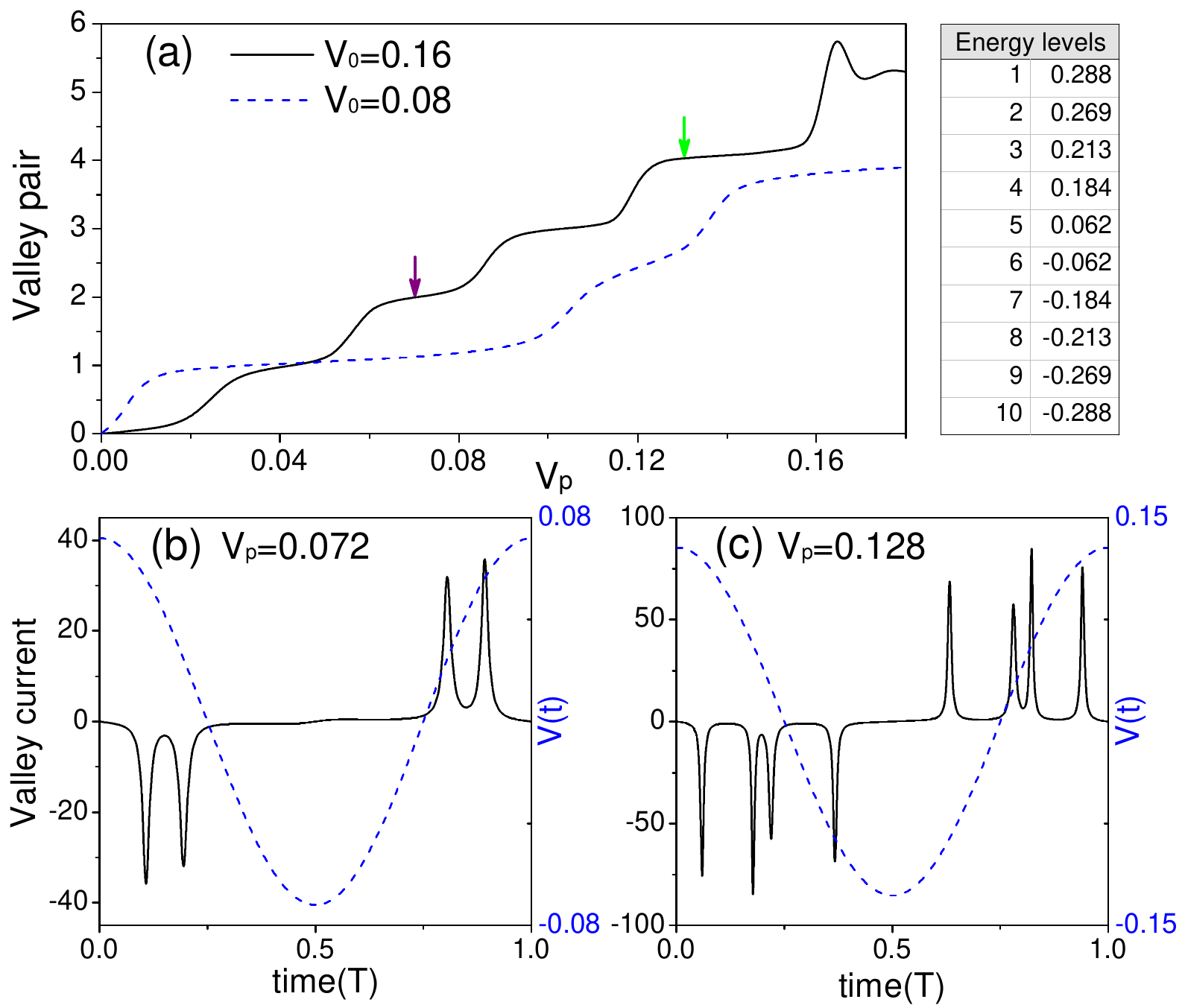}
\caption{(a) Number of valley pairs as a function of $V_p$ for two static potentials. The table lists the energy levels near zero energy for an isolated SQD with $V_0=0.16$, $W=3$, and $H=20$. (b) and (c) Instantaneous valley current over one period, $T=2\pi/\omega$, for two values of $V_p$. The blue dashed curves show the periodic potential $V(t)=V_p\cos(\omega t)$.}\label{fig3}
\end{figure}

We next examine the influence of the static potential $V_0$, which shifts the QD energy levels. Figure~\ref{fig3}(a) shows the number of valley pairs pumped by an SQD device as a function of $V_p$ for two values of $V_0$. The static potential produces additional quantized plateaus. For $V_0=0.08$, a broad plateau occurs at one valley pair. For $V_0=0.16$, the response develops a staircase of integer plateaus from one to four valley pairs, resembling conductance quantization in mesoscopic transport. To explain this staircase, the table in Fig.~\ref{fig3}(a) lists the energy levels near zero energy for an isolated SQD at $V_0=0.16$, while Figs.~\ref{fig3}(b) and \ref{fig3}(c) show the instantaneous valley currents over one period. The values $V_p=0.072$ and $V_p=0.128$, indicated by the blue and purple arrows in Fig.~\ref{fig3}(a), correspond to the second and fourth plateaus, respectively. The tabulated levels include the static potential, whereas the periodic potential shown in Figs.~\ref{fig3}(b) and \ref{fig3}(c) is $V(t)=V_p\cos(\omega t)$. The instantaneous valley current consists of resonant peaks, and the number of peaks in half a cycle agrees with the integer value of the corresponding plateau. These peaks occur when $V(t)$ sweeps across the discrete SQD energy levels. In Fig.~\ref{fig3}(b), the periodic potential crosses two levels and produces two current peaks in half a cycle. For $V_p=0.128$, four peaks appear in Fig.~\ref{fig3}(c), with two closely spaced peaks separated from the other pair.

The instantaneous gate potentials $V(t)+V_0$ at the two current peaks in Fig.~\ref{fig3}(b) are $0.1844$ and $0.2163$ eV, close to the fourth and third QD energy levels in the table, $0.184$ and $0.213$ eV, respectively. In Fig.~\ref{fig3}(c), the corresponding values are $0.0742$, $0.1848$, $0.2167$, and $0.2793$ eV, which approximately match the fifth through second levels, $0.062$, $0.184$, $0.213$, and $0.269$ eV, respectively. This correspondence supports the interpretation that the current peaks arise from resonant transport through discrete QD levels.

\begin{figure}[tbp]
\centering
\includegraphics[width=0.9\columnwidth]{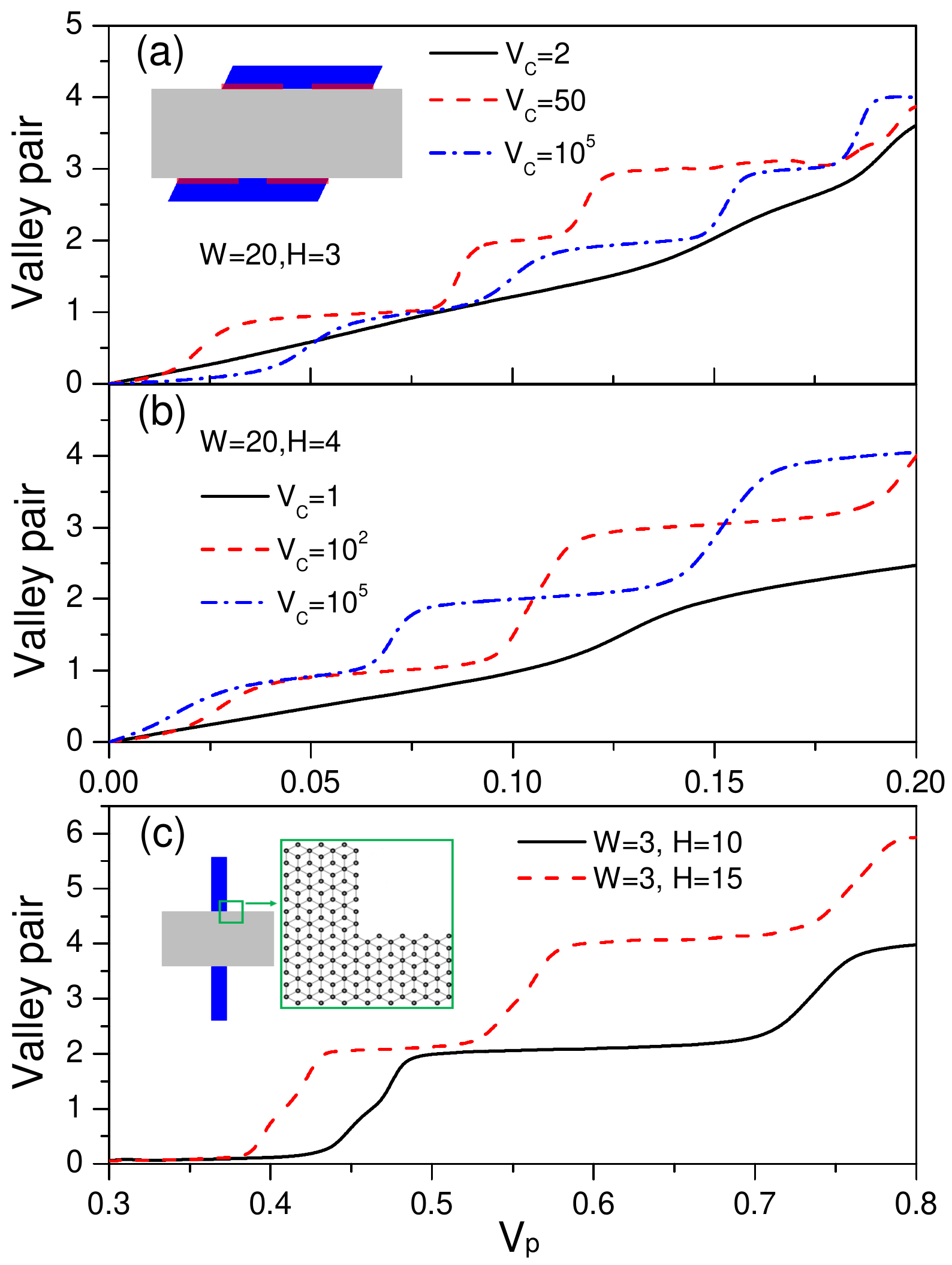}
\caption{(a) and (b) Number of valley pairs as a function of the pumping amplitude $V_p$ for several values of the confinement potential $V_C$. (c) Number of valley pairs as a function of $V_p$ for SQDs with armchair boundaries.}\label{fig4}
\end{figure}

We attribute the approximately linear response of the FQD devices to strong coupling between the QDs and the bilayer graphene ribbon. To verify this point, we introduce a confinement potential $V_C$ at the QD-ribbon interface, as illustrated in the inset of Fig.~\ref{fig4}(a). For QD heights $H=3$ and $H=4$, Figs.~\ref{fig4}(a) and \ref{fig4}(b) show the number of valley pairs as a function of $V_p$ for several values of $V_C$. Increasing $V_C$ weakens the QD-ribbon coupling and suppresses the approximately linear response. At sufficiently large $V_C$, quantized plateaus emerge, as shown by the red dashed and blue dash-dotted curves. Thus, reducing the coupling between an FQD and the ribbon restores quantized valley emission. The plateau values can also be tuned through $V_C$, allowing the number of valley pairs emitted into the leads to be controlled by the interface potential. These results show that the plateau structure is determined jointly by the QD energy-level distribution and the strength of the QD-ribbon coupling.

So far we have considered QDs with zigzag boundaries. We now examine whether quantized valley pumping can occur for armchair QDs by coupling two armchair SQDs to the bilayer graphene ribbon. The configuration is illustrated in the inset of Fig.~\ref{fig4}(c). Figure~\ref{fig4}(c) demonstrates the number of valley pairs as a function of $V_p$ for a fixed width $W=3$ and two heights. For both $H=10$ and $H=15$, the number of valley pairs accumulated in half a cycle forms quantized plateaus at $2$, $4$, and higher even integers. Compared with zigzag QDs, the different energy-level spectrum of the armchair QDs shifts the plateaus to larger gate voltages. The increased spacing between the relevant QD levels produces the well-defined quantized plateaus shown in Fig.~\ref{fig4}(c).

\begin{figure}[tbp]
\centering
\includegraphics[width=0.9\columnwidth]{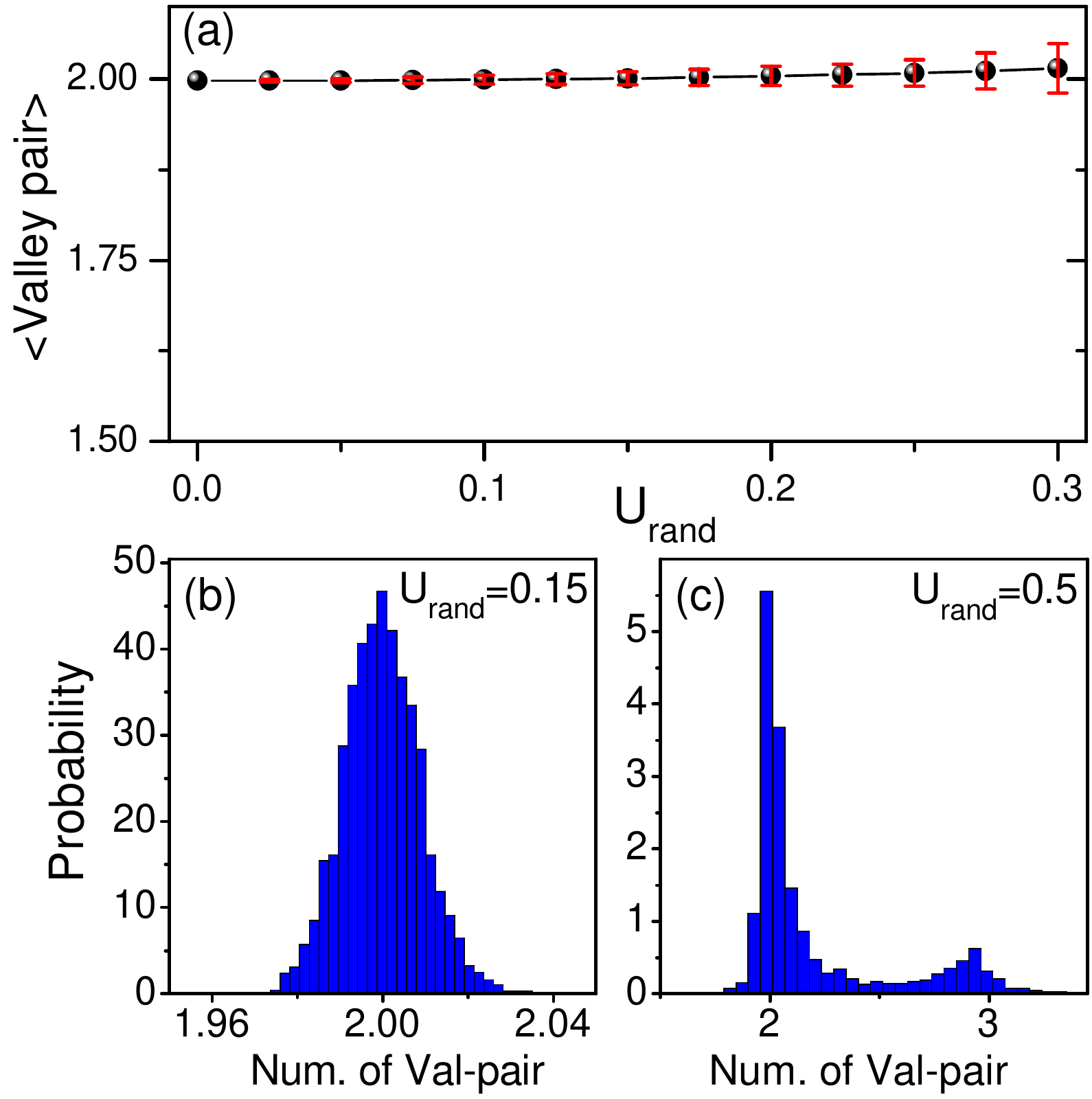}
\caption{(a) Disorder-averaged number of valley pairs as a function of the disorder strength $U_{rand}$. The red error bars denote the standard deviation over disorder configurations. (b) and (c) Distributions of the number of valley pairs for $U_{rand}=0.15$ and $U_{rand}=0.5$, respectively. The QD dimensions are $W=3$ and $H=20$, the central ribbon has a width of $80$, and the pumping amplitude is $V_p=0.144$.}\label{fig5}
\end{figure}

Finally, we examine the effect of disorder confined to the QDs. Anderson disorder is introduced as an independent random on-site potential at each QD site, distributed uniformly in $[-U_{rand}/2,U_{rand}/2]$, where $U_{rand}$ is the disorder strength. Figure~\ref{fig5} presents results for SQDs with $W=3$ and $H=20$, as in Fig.~\ref{fig2}(a), with the parameters chosen on the first quantized plateau. As shown in Fig.~\ref{fig5}(a), disorder has little effect on the averaged number of valley pairs, which remains close to the quantized value of $2$ with small fluctuations. Counterintuitively, the average pair number increases slightly with $U_{rand}$. To clarify this behavior, Figs.~\ref{fig5}(b) and \ref{fig5}(c) show distributions accumulated over $3,000$ random configurations. At $U_{rand}=0.15$, the distribution is narrow and almost symmetric about $2$, yielding an average of two valley pairs. At $U_{rand}=0.5$, the distribution develops two peaks near $2$ and $3$, thereby increasing the average pair number. This disorder-enhanced pumping can be understood from the evolution of the QD spectrum. Random on-site potentials mix and shift the QD energy levels, modifying both the level-spacing distribution and the spectral weight of discrete resonances. Such disorder-induced changes in spectral correlations are well established for mesoscopic systems\cite{Mirlin2000Statistics}, while calculations for graphene QDs show explicitly that bulk and edge disorder reshape the density of bound states and the associated transport resonances\cite{Pal2011Disorder}. In the present calculations, increasing $U_{rand}$ reduces the mean spacing between the QD levels relevant to the pumping process. Consequently, an additional QD resonance enters the energy window swept by the periodic gate for a growing fraction of disorder configurations. These configurations emit three rather than two valley pairs, accounting for both the second peak near $3$ in Fig.~\ref{fig5}(c) and the gradual increase in the average pair number in Fig.~\ref{fig5}(a). This mechanism is also consistent with the static-potential dependence shown in Fig.~\ref{fig3}(a), where a change in the QD spectrum shifts the pumping plateau.

\section{Conclusion}

In summary, we propose a quantized valley source based on adiabatic quantum pumping through two QDs coupled to a bilayer graphene ribbon with coexisting QSH and QVH effects. Changing the relative phase of the two gate voltages switches between a pure charge pump and a pure valley pump. Under out-of-phase pumping, slender QDs emit an integer number of valley pairs in half a cycle. The quantized valley pairs arise from discrete QD levels entering the pumping window. A static gate shifts these levels, while the QD shape and its coupling to the ribbon determine whether the pumped valley pairs can be quantized as plateaus. This explains why strongly coupled flattened QDs show a continuous response and why quantized plateaus emerge when the coupling is reduced. The quantized phenomenon is found for both zigzag and armchair QD boundaries. It is also robust against Anderson disorder confined to the QDs. These results establish topological edge-state selectivity and adiabatic quantum pumping as complementary mechanisms for generating a quantized valley source.

\section*{Acknowledgments}
This work was supported by the National Natural Science Foundation of China (Grants No.~12574054, No.~12304061, and No.~12474047). F.X. also acknowledges support from the Shenzhen Science and Technology Program (Grant No.~ZDYJ20251211120859002).


\begin{thebibliography}{00}
\bibitem{LundebergHarnessing}
M. B. Lundeberg and J. A. Folk, Harnessing chirality for valleytronics, \href{https://doi.org/10.1126/science.1260989}{Science \textbf{346}, 422 (2014).}

\bibitem{Schaibley2016Valleytronics}
J. R. Schaibley, H. Yu, G. Clark, P. Rivera, J. S. Ross, K. L. Seyler, W. Yao, and X. Xu, Valleytronics in 2D materials, \href{https://doi.org/10.1038/natrevmats.2016.55}{Nat. Rev. Mater. \textbf{1}, 16055 (2016).}

\bibitem{HuMoire}
C. Hu, V. Michaud-Rioux, W. Yao, and H. Guo, Moir\'{e} Valleytronics: Realizing Dense Arrays of Topological Helical Channels, \href{https://doi.org/10.1103/physrevlett.121.186403}{Phys. Rev. Lett. \textbf{121}, 186403 (2018).}

\bibitem{StevenValleytronics}
S. A. Vitale, D. Nezich, J. O. Varghese, P. Kim, N. Gedik, P. Jarillo-Herrero, D. Xiao, and M. Rothschild, Valleytronics: Opportunities, Challenges, and Paths Forward, \href{https://doi.org/10.1002/smll.201801483}{Small \textbf{14}, 1801483 (2018).}

\bibitem{Yanping2019Valleytronics}
Y. Liu, Y. Gao, S. Zhang, J. He, J. Yu, and Z. Liu, Valleytronics in transition metal dichalcogenides materials, \href{https://doi.org/10.1007/s12274-019-2497-2}{Nano Res. \textbf{12}, 2695 (2019).}

\bibitem{ChenValley}
X. Chen, Y. Xu, J. Wang, and H. Guo, Valley filtering effect of phonons in graphene with a grain boundary, \href{https://doi.org/10.1103/physrevb.99.064302}{Phys. Rev. B \textbf{99}, 064302 (2019).}

\bibitem{Zhi2020Valley}
Z.-M. Yu, S. Guan, X.-L. Sheng, W. Gao, and S. A. Yang, Valley-Layer Coupling: A New Design Principle for Valleytronics, \href{https://doi.org/10.1103/physrevlett.124.037701}{Phys. Rev. Lett. \textbf{124}, 037701 (2020).}

\bibitem{MakControl}
K. F. Mak, K. He, J. Shan, and T. F. Heinz, Control of valley polarization in monolayer ${\mathrm{MoS}}_{2}$ by optical helicity, \href{https://doi.org/10.1038/nnano.2012.96}{Nat. Nanotechnol. \textbf{7}, 494 (2012).}

\bibitem{Zhang14Generation}
L. Zhang, K. Gong, J. Chen, L. Liu, Y. Zhu, D. Xiao, and H. Guo, Generation and transport of valley-polarized current in transition-metal dichalcogenides, \href{https://doi.org/10.1103/PhysRevB.90.195428}{Phys. Rev. B \textbf{90}, 195428 (2014).}

\bibitem{Ye2016Optical}
Z. Ye, D. Sun, and T. F. Heinz, Optical manipulation of valley pseudospin, \href{https://doi.org/10.1038/nphys3891}{Nat. Phys. \textbf{13}, 26 (2017).}

\bibitem{Yang24Photocurrent}
Y. Yang, X. Cheng, L. Xiao, S. Jia, J. Chen, L. Zhang, and J. Wang, Electric field controlled valley-polarized photocurrent switch based on the circular bulk photovoltaic effect, \href{https://doi.org/10.1103/PhysRevB.109.235403}{Phys. Rev. B \textbf{109}, 235403 (2024).}

\bibitem{Sharma24Direct}
S. Sharma, D. Gill, J. Krishna, J. K. Dewhurst, and S. Shallcross, Direct coupling of light to valley current, \href{https://doi.org/10.1038/s41467-024-51968-5}{Nat. Commun. \textbf{15}, 7579 (2024).}

\bibitem{Gill25Creation}
D. Gill, S. Sharma, J. K. Dewhurst, and S. Shallcross, Creation and control of valley currents in graphene by few cycle light pulses, \href{https://doi.org/10.1038/s41524-025-01689-0}{npj Comput. Mater. \textbf{11}, 185 (2025).}

\bibitem{GunlyckeGraphene} 
D. Gunlycke and C. T. White, Graphene Valley Filter Using a Line Defect, \href{https://doi.org/10.1103/physrevlett.106.136806}{Phys. Rev. Lett. \textbf{106}, 136806 (2011).}

\bibitem{Fuming2016Transmission}
F. Xu, Z. Yu, Y. Ren, B. Wang, Y. Wei, and Z. Qiao, Transmission spectra and valley processing of graphene and carbon nanotube superlattices with inter-valley coupling, \href{https://doi.org/10.1088/1367-2630/18/11/113011}{New J. Phys. \textbf{18}, 113011 (2016).}

\bibitem{LiValley}
S.-Y. Li, Y. Su, Y.-N. Ren, and L. He, Valley Polarization and Inversion in Strained Graphene via Pseudo-Landau Levels, Valley Splitting of Real Landau Levels, and Confined States, \href{https://doi.org/10.1103/physrevlett.124.106802}{Phys. Rev. Lett. \textbf{124}, 106802 (2020).}

\bibitem{Park2019Magnetoelectrically}
C. Park, Magnetoelectrically Controlled Valley Filter and Valley Valve in Bilayer Graphene, \href{https://doi.org/10.1103/physrevapplied.11.044033}{Phys. Rev. Appl. \textbf{11}, 044033 (2019).}

\bibitem{CaiMagnetic}
T. Cai, S. A. Yang, X. Li, F. Zhang, J. Shi, W. Yao, and Q. Niu, Magnetic control of the valley degree of freedom of massive Dirac fermions with application to transition metal dichalcogenides, \href{https://doi.org/10.1103/physrevb.88.115140}{Phys. Rev. B \textbf{88}, 115140 (2013).}

\bibitem{AnRealization} 
X.-T. An, J. Xiao, M. W.-Y. Tu, H. Yu, V. I. Fal'ko, and W. Yao, Realization of Valley and Spin Pumps by Scattering at Nonmagnetic Disorders, \href{https://doi.org/10.1103/physrevlett.118.096602}{Phys. Rev. Lett. \textbf{118}, 096602 (2017).}

\bibitem{An20Klein}
X.-T. An and W. Yao, Valley-selective Klein tunneling through a superlattice barrier in graphene, \href{https://doi.org/10.1103/PhysRevApplied.14.014039}{Phys. Rev. Applied \textbf{14}, 014039 (2020).}

\bibitem{Zhang23Electrical}
Q.-Q. Zhang, X.-T. An, and J.-J. Liu, Electrical tuning of valley polarization in monolayer transition metal dichalcogenides, \href{https://doi.org/10.1103/PhysRevB.107.035411}{Phys. Rev. B \textbf{107}, 035411 (2023).}

\bibitem{Mu22Fano}
H.-Y. Mu, N.-W. Wang, Y.-N. Du, X.-T. An, and J.-J. Liu, Valley-resolved Fano resonance in monolayer transition metal dichalcogenide nanoribbons with attached stubs, \href{https://doi.org/10.1103/PhysRevB.105.115305}{Phys. Rev. B \textbf{105}, 115305 (2022).}

\bibitem{Shao24NonHermitian}
K. Shao, H. Geng, E. Liu, J. L. Lado, W. Chen, and D. Y. Xing, Non-Hermitian moir\'{e} valley filter, \href{https://doi.org/10.1103/PhysRevLett.132.156301}{Phys. Rev. Lett. \textbf{132}, 156301 (2024).}

\bibitem{Yan25Giant}
K. Yan, L. Cheng, Y. Hu, J. Gao, X. Zou, and X. Chen, Giant tunneling magnetoresistance based on spin-valley-mismatched ferromagnetic metals, \href{https://doi.org/10.1103/PhysRevLett.134.036302}{Phys. Rev. Lett. \textbf{134}, 036302 (2025).}

\bibitem{AngValley} 
Y. S. Ang, S. A. Yang, C. Zhang, Z. Ma, and L. K. Ang, Valleytronics in merging Dirac cones: All-electric-controlled valley filter, valve, and universal reversible logic gate, \href{https://doi.org/10.1103/physrevb.96.245410}{Phys. Rev. B \textbf{96}, 245410 (2017).}

\bibitem{FeveAn} 
G.~F\`{e}ve, A. Mah\'{e}, J.-M. Berroir, T. Kontos, B. Placais, D. C. Glattli, A. Cavanna, B. Etienne, and Y. Jin, An on-demand coherent single-electron source, \href{https://doi.org/10.1126/science.1141243}{Science \textbf{316}, 1169 (2007).}

\bibitem{MoskaletsQuantized} 
M. Moskalets, P. Samuelsson, and M. B{\"u}ttiker, Quantized Dynamics of a Coherent Capacitor, \href{https://doi.org/10.1103/physrevlett.100.086601}{Phys. Rev. Lett. \textbf{100}, 086601 (2008).}

\bibitem{Gurzhi2003Spin}
R. N. Gurzhi, A. N. Kalinenko, A. I. Kopeliovich, A. V. Yanovsky, E. N. Bogachek, and U. Landman, Spin-guide source for the generation of highly spin-polarized currents, \href{https://doi.org/10.1103/physrevb.68.125113}{Phys. Rev. B \textbf{68}, 125113 (2003).}

\bibitem{Xing2014Coherent}
Y. Xing, Z. L. Yang, Q.-f. Sun, and J. Wang, Coherent single-spin source based on topological insulators, \href{https://doi.org/10.1103/physrevb.90.075435}{Phys. Rev. B \textbf{90}, 075435 (2014).}

\bibitem{XiaoCoupled}
D. Xiao, G.-B. Liu, W. Feng, X. Xu, and W. Yao, Coupled Spin and Valley Physics in Monolayers of ${\mathrm{MoS}}_{2}$ and Other Group-VI Dichalcogenides, \href{https://doi.org/10.1103/physrevlett.108.196802}{Phys. Rev. Lett. \textbf{108}, 196802 (2012).}

\bibitem{KhaniGate}
H. Khani and S. Piri Pishekloo, Gate-controlled spin-valley-layer locking in bilayer transition-metal dichalcogenides, \href{https://doi.org/10.1039/d0nr04630b}{Nanoscale \textbf{12}, 22281 (2020).}

\bibitem{EzawaSpin}
M. Ezawa, Spin valleytronics in silicene: Quantum spin Hall-quantum anomalous Hall insulators and single-valley semimetals, \href{https://doi.org/10.1103/physrevb.87.155415}{Phys. Rev. B \textbf{87}, 155415 (2013).}

\bibitem{XiongSpin}
G. Xiong, D. Wang, F. Xu, Y. Ma, X. Li, T. Wei, and Y. Wang, Spin-valley polarized transport in a field-controllable bilayer silicene superlattice, \href{https://doi.org/10.1103/physrevb.106.205409}{Phys. Rev. B \textbf{106}, 205409 (2022).}

\bibitem{LiuSpin2021}
J. Y. Liu, J. Yu, J. L. Ning, H. M. Yi, L. Miao, L. J. Min, Y. F. Zhao, W. Ning, K. A. Lopez, Y. L. Zhu, T. Pillsbury, Y. B. Zhang, Y. Wang, J. Hu, H. B. Cao, B. C. Chakoumakos, F. Balakirev, F. Weickert, M. Jaime, Y. Lai, K. Yang, J. W. Sun, N. Alem, V. Gopalan, C. Z. Chang, N. Samarth, C. X. Liu, R. D. McDonald, and Z. Q. Mao, Spin-valley locking and bulk quantum Hall effect in a noncentrosymmetric Dirac semimetal ${\mathrm{BaMnSb}}_{2}$, \href{https://doi.org/10.1038/s41467-021-24369-1}{Nat. Commun. \textbf{12}, 4062 (2021).}

\bibitem{Ominato2020}
Y. Ominato, J. Fujimoto, and M. Matsuo, Valley-Dependent Spin Transport in Monolayer Transition-Metal Dichalcogenides, \href{https://doi.org/10.1103/physrevlett.124.166803}{Phys. Rev. Lett. \textbf{124}, 166803 (2020).}

\bibitem{Yang2020}
J.-E. Yang, X.-L. L{\"u}, C.-X. Zhang, and H. Xie, Topological spin-valley filtering effects based on hybrid silicene-like nanoribbons, \href{https://doi.org/10.1088/1367-2630/ab84b4}{New J. Phys. \textbf{22}, 053034 (2020).}

\bibitem{wang2023}
K.-T. Wang, H. Wang, F. Xu, Y. Yu, and Y. Wei, Realization of valley-spin polarized current via parametric pump in monolayer ${\mathrm{MoS}}_{2}$, \href{https://doi.org/10.1088/1367-2630/acb22e}{New J. Phys. \textbf{25}, 013019 (2023).}

\bibitem{Wang24Electrical}
T. Wang, M. Vila, M. P. Zaletel, and S. Chatterjee, Electrical control of spin and valley in spin-orbit-coupled graphene multilayers, \href{https://doi.org/10.1103/PhysRevLett.132.116504}{Phys. Rev. Lett. \textbf{132}, 116504 (2024).}

\bibitem{ZhuQuantum} 
W. Zhu, Y. Long, H. Chen, and J. Ren, Quantum valley Hall effects and spin-valley locking in topological Kane-Mele circuit networks, \href{https://doi.org/10.1103/physrevb.99.115410}{Phys. Rev. B \textbf{99}, 115410 (2019).}

\bibitem{TaoTwo} 
L. L. Tao and E. Y. Tsymbal, Two-dimensional spin-valley locking spin valve, \href{https://doi.org/10.1103/physrevb.100.161110}{Phys. Rev. B \textbf{100}, 161110(R) (2019).}

\bibitem{TaoValley} 
L. L. Tao, A. Naeemi, and E. Y. Tsymbal, Valley-Spin Logic Gates, \href{https://doi.org/10.1103/physrevapplied.13.054043}{Phys. Rev. Appl. \textbf{13}, 054043 (2020).}

\bibitem{WuSpin} 
Y. Wu, Q. Tong, G.-B. Liu, H. Yu, and W. Yao, Spin-valley qubit in nanostructures of monolayer semiconductors: Optical control and hyperfine interaction, \href{https://doi.org/10.1103/physrevb.93.045313}{Phys. Rev. B \textbf{93}, 045313 (2016).}

\bibitem{GImpurity} 
G. Sz\'{e}chenyi, L. Chirolli, and A. P\'{a}lyi, Impurity-assisted electric control of spin-valley qubits in monolayer ${\mathrm{MoS}}_{2}$, \href{https://doi.org/10.1088/2053-1583/aab80e}{2D Mater. \textbf{5}, 035004 (2018).}

\bibitem{PawlowskiSpin} 
J. Paw\l{}owski, Spin-valley system in a gated ${\mathrm{MoS}}_{2}$-monolayer quantum dot, \href{https://doi.org/10.1088/1367-2630/ab5ac9}{New J. Phys. \textbf{21}, 123029 (2019).}

\bibitem{BernevigQuantum} 
B. A. Bernevig, T. L. Hughes, and S.-C. Zhang, Quantum spin Hall effect and topological phase transition in HgTe quantum wells, \href{https://doi.org/10.1126/science.1133734}{Science \textbf{314}, 1757 (2006).}

\bibitem{KonigQuantum} 
M. K{\"o}nig, S. Wiedmann, C. Br{\"u}ne, A. Roth, H. Buhmann, L. W. Molenkamp, X.-L. Qi, and S.-C. Zhang, Quantum spin Hall insulator state in HgTe quantum wells, \href{https://doi.org/10.1126/science.1148047}{Science \textbf{318}, 766 (2007).}

\bibitem{ZhangTopological} 
H. Zhang, C.-X. Liu, X.-L. Qi, X. Dai, Z. Fang, and S.-C. Zhang, Topological insulators in ${\mathrm{Bi}}_{2}{\mathrm{Se}}_{3}$, ${\mathrm{Bi}}_{2}{\mathrm{Te}}_{3}$ and ${\mathrm{Sb}}_{2}{\mathrm{Te}}_{3}$ with a single Dirac cone on the surface, \href{https://doi.org/10.1038/nphys1270}{Nat. Phys. \textbf{5}, 438 (2009).}

\bibitem{ChenExperimental} 
Y. L. Chen, J. G. Analytis, J.-H. Chu, Z. K. Liu, S.-K. Mo, X.-L. Qi, H. J. Zhang, D. H. Lu, X. Dai, Z. Fang, S.-C. Zhang, I. R. Fisher, Z. Hussain, Z.-X. Shen, Experimental Realization of a Three-Dimensional Topological Insulator, ${\mathrm{Bi}}_{2}{\mathrm{Te}}_{3}$, \href{https://doi.org/10.1126/science.1173034}{Science \textbf{325}, 178 (2009).}

\bibitem{Wu2017Observation} 
S. Wu, V. Fatemi, Q. D. Gibson, K. Watanabe, T. Taniguchi, R. J. Cava, and P. Jarillo-Herrero, Observation of the quantum spin Hall effect up to 100 kelvin in a monolayer crystal, \href{https://doi.org/10.1126/science.aan6003}{Science \textbf{359}, 76 (2018).}

\bibitem{VergnioryA} 
M. G. Vergniory, L. Elcoro, C. Felser, N. Regnault, B. A. Bernevig, and Z. Wang, A complete catalogue of high-quality topological materials, \href{https://doi.org/10.1038/s41586-019-0954-4}{Nature \textbf{566}, 480 (2019).}

\bibitem{LiuQuantum}
C.-C. Liu, W. Feng, and Y. Yao, Quantum Spin Hall Effect in Silicene and Two-Dimensional Germanium, \href{https://doi.org/10.1103/physrevlett.107.076802}{Phys. Rev. Lett. \textbf{107}, 076802 (2011).}

\bibitem{XiaoValley}
D. Xiao, W. Yao, and Q. Niu, Valley-Contrasting Physics in Graphene: Magnetic Moment and Topological Transport, \href{https://doi.org/10.1103/physrevlett.99.236809}{Phys. Rev. Lett. \textbf{99}, 236809 (2007).}

\bibitem{YuValley}
Z. Yu, F. Xu, and J. Wang, Valley Seebeck effect in gate tunable zigzag graphene nanoribbons, \href{https://doi.org/10.1016/j.carbon.2015.12.033}{Carbon \textbf{99}, 451 (2016).}

\bibitem{ZhangInfluence}
L. Zhang, Z. Yu, F. Xu, and J. Wang, Influence of dephasing and B/N doping on valley Seebeck effect in zigzag graphene nanoribbons, \href{https://doi.org/10.1016/j.carbon.2017.10.017}{Carbon \textbf{126}, 183 (2018).}

\bibitem{Pan2014Valley}
H. Pan, Z. Li, C.-C. Liu, G. Zhu, Z. Qiao, and Y. Yao, Valley-Polarized Quantum Anomalous Hall Effect in Silicene, \href{https://doi.org/10.1103/physrevlett.112.106802}{Phys. Rev. Lett. \textbf{112}, 106802 (2014).}

\bibitem{Pan2015Valley}
H. Pan, X. Li, H. Jiang, Y. Yao, and S. A. Yang, Valley-polarized quantum anomalous Hall phase and disorder-induced valley-filtered chiral edge channels, \href{https://doi.org/10.1103/physrevb.91.045404}{Phys. Rev. B \textbf{91}, 045404 (2015).}

\bibitem{Zhai2020}
X. Zhai and Y. M. Blanter, Spin-valley polarized quantum anomalous Hall effect and a valley-controlled half-metal in bilayer graphene, \href{https://doi.org/10.1103/physrevb.101.155425}{Phys. Rev. B \textbf{101}, 155425 (2020).}

\bibitem{Kazantsev24Nonconservation}
A. Kazantsev, A. Mills, E. O'Neill, H. Sun, G. Vignale, and A. Principi, Nonconservation of the valley density and its implications for the observation of the valley Hall effect, \href{https://doi.org/10.1103/PhysRevLett.132.106301}{Phys. Rev. Lett. \textbf{132}, 106301 (2024).}

\bibitem{QiaoTwo}
Z. Qiao, W.-K. Tse, H. Jiang, Y. Yao, and Q. Niu, Two-Dimensional Topological Insulator State and Topological Phase Transition in Bilayer Graphene, \href{https://doi.org/10.1103/physrevlett.107.256801}{Phys. Rev. Lett. \textbf{107}, 256801 (2011).}

\bibitem{Belayadi25Dynamic}
A. Belayadi, C. I. Osuala, I. Assi, N. A. Hadadi, J. P. F. LeBlanc, and A. Abbout, Graphene/hBN-based valley transistor: Dynamic control of valley current in synchronized nonzero voltages within the time-dependent regime, \href{https://doi.org/10.1103/ns2x-rfcf}{Phys. Rev. B \textbf{112}, 165411 (2025).}

\bibitem{SwitkesAn}
M. Switkes, C. M. Marcus, K. Campman, and A. C. Gossard, An adiabatic quantum electron pump, \href{https://doi.org/10.1126/science.283.5409.1905}{Science \textbf{283}, 1905 (1999).}

\bibitem{WeiResonance}
Y. Wei, J. Wang, and H. Guo, Resonance-assisted parametric electron pump, \href{https://doi.org/10.1103/physrevb.62.9947}{Phys. Rev. B \textbf{62}, 9947 (2000).}

\bibitem{WeiCarbon}
Y. Wei and J. Wang, Carbon-nanotube-based quantum pump in the presence of a superconducting lead, \href{https://doi.org/10.1103/physrevb.66.195419}{Phys. Rev. B \textbf{66}, 195419 (2002).}

\bibitem{WangCurrent}
B. Wang, J. Wang, and H. Guo, Current plateaus of nonadiabatic charge pump: multiphoton assisted processes, \href{https://doi.org/10.1103/physrevb.68.155326}{Phys. Rev. B \textbf{68}, 155326 (2003).}

\bibitem{XuNumerical}
F. Xu, Y. Xing, and J. Wang, Numerical study of parametric pumping current in mesoscopic systems in the presence of a magnetic field, \href{https://doi.org/10.1103/physrevb.84.245323}{Phys. Rev. B \textbf{84}, 245323 (2011).}

\bibitem{XingTopological}
Y. Xing, L. Zhang, and J. Wang, Topological Anderson insulator phenomena, \href{https://doi.org/10.1103/physrevb.84.035110}{Phys. Rev. B \textbf{84}, 035110 (2011).}

\bibitem{WangRelation}
J. Wang and H. Guo, Relation between nonequilibrium Green's function and Lippmann-Schwinger formalism in the first-principles quantum transport theory, \href{https://doi.org/10.1103/physrevb.79.045119}{Phys. Rev. B \textbf{79}, 045119 (2009).}

\bibitem{Yu2008Rashba} 
Y. S. Dedkov, M. Fonin, U. R{\"u}diger, and C. Laubschat, Rashba Effect in the Graphene/Ni(111) System, \href{https://doi.org/10.1103/physrevlett.100.107602}{Phys. Rev. Lett. \textbf{100}, 107602 (2008).}

\bibitem{Mirlin2000Statistics}
A. D. Mirlin, Statistics of energy levels and eigenfunctions in disordered systems, \href{https://doi.org/10.1016/S0370-1573(99)00091-5}{Phys. Rep. \textbf{326}, 259 (2000).}

\bibitem{Pal2011Disorder}
G. Pal, W. Apel, and L. Schweitzer, Electric transport through circular graphene quantum dots: Presence of disorder, \href{https://doi.org/10.1103/PhysRevB.84.075446}{Phys. Rev. B \textbf{84}, 075446 (2011).}

\end{thebibliography}
\end{document}